\documentstyle[11pt]{article}
\date{}
\begin{document}
\sloppy
\title{The Force Exerting on Cosmic Bodies in a Quaternionic
Field}
\author{V. Majern\'{\i}k\\
Department of Theoretical Physics,
Palack\'y University,\\ T\v r. 17. Listopadu 50, CZ-772 07 Olomouc,
Czech Republic\\ and\\
Institute of Mathematics, Slovak
Academy of Sciences, \\ Bratislava, \v Stef\'anikova  47,
 Slovak Republic}
\maketitle
\begin{abstract}
The expression of a time-dependent
cosmological constant $\lambda \propto 1/t^2$ is interpreted as
the energy density of a special type of the quaternionic field.
The Lorenz-like force acting on the
moving body in the presence of this quaternionic field is determined.
The astronomical and
terrestrial effects of this field are presented, and the ways how it can be
observably detected is discussed. Finally, a new mechanism of the particle
creation and an alternative cosmological scenario in the presence of the
cosmic quatertionic field is suggested.

\end{abstract}
%--------------------------------------------------------------------------\\
\section{Introduction}
%-----------------------------------------------------------------------------\\

As is well-known one way to account for the possible cosmic acceleration
is the introduction a new type
of energy, the so-called {\it quintessence} ("dark energy"), a dynamical,
spatially inhomogeneous form of energy with negative pressure \cite{ST}.
A common example is the energy of a slowly evolving scalar field with positive
potential energy, similar to the inflation field in the inflation
cosmology.
The quintessence cosmological scenario (QCDM) is a
spatially flat FRW space-time dominated by the radiation at early times,
and cold dark matter (CDM) and quintessence (Q) later time.
The quintessence is supposed to obey an equation of state of the form
\begin{equation} \label{1}
p_Qc^{-2}=w_Q\varrho_Q,\qquad
-1<w_Q<0.
\end{equation}
In many models $w_Q$ can vary over time. For the vacuum energy (static
cosmological
constant), it holds $w_Q=-1$ and $\dot w_Q=0$.
The existence of the quintessence, often modelled
by a positive non-zero  cosmological constant, helps
to overcome the cosmological age and tuning problems.
The point of view has often been adopted
which allows the quintessence to vary in time, i.e. $\omega_Q=f(t)$.
This means that the corresponding cosmical constant is time-dependent, too.
Next, we will
consider a cosmological constant $\lambda \propto 1/t^2$.

As is well-known the Einstein field equations with a non-zero $\lambda$
can be rearranged so
that their right-hand sides
have two terms: the stress-energy tensor of the ordinary
matter and an additional stress-energy tensor $T_{ij}^{(v)}$
assigned to $\lambda$
$$T^{(\nu)}_{ij}=\Lambda=\left (\frac{c^4\lambda}{8\pi G}\right ).$$
A number of authors set phenomenologically
$\Lambda\propto 1/t^2$ [3-10]
%\cite{EF} \cite{B} \cite{BS} \cite{L} \cite{BE}
%\cite{LN} \cite{AL} \cite{ARB}
(for a review see \cite{OO}).
Generally, $\Lambda$  contains in its definition the gravitation
constant $G$ and velocity of light $c$.
The simplest expression for $\Lambda\propto 1/t^2$, having the right
dimension,
and containing $G$ and $c$ is
$$\Lambda=\frac{c^2}{8\pi Gt^2},$$
where $\kappa$ is a dimensionless constant.

In a very recent article \cite{MK},
$\Lambda$ has been interpreted as the {\it field energy} of a  classical
quaternionic field (called $\Phi$-field, for short)
by written it in the form
$$\Lambda=\frac{1}{8\pi }\left [\frac{c}{\sqrt{G}t}
\frac{c}{\sqrt{G}t}\right ]
=\frac{\Phi^{2}}{8\pi}\qquad \Phi=\frac{c}{\sqrt{G}t},\quad \eqno(1)
$$
where $\Phi$ is the intensity of a special quaternionic field \cite{MM}
\cite{SIG} \cite{A} which is
given by the field tensor $F_{ij}\quad i,j=1,2,3,0$ whose
components are defined as
$F_{ij}=0$ for $i \neq j$ and $F_{11}=F_{22}=F_{33}=-F_{00}=\Phi.$
The $\Phi$-field belongs to the family of the quaternionic
fields (see \cite{MM}).
The quaternionic field which we consider is given
by the field tensor which, in the matrix, has the form
$$F_{ij}=
\left(\begin{array}{cccc}
\Phi&0&0&0\\
0&\Phi&0&0\\
0&0&\Phi&0\\
0&0&0&-\Phi
\end{array} \right).$$
$\Phi$ is the only field variable in it.
$F_{ij}$ is a symmetric field tensor with
the components $F_{ii}=\Phi\quad i=1,2,3,\quad F_{ii}=-\Phi\quad
i=0,$ and $F_{ij}=0\quad
i\neq j.$  It is easily to show that $\Phi$ is transformed as a scalar
under Lorentz
transformation \cite{MK}.
The field equations of the $\Phi$-field in the differential are
$$\nabla \Phi=k \vec J\quad\quad i=1,2,3\quad\quad
{\rm and}\qquad
-{1\over c}{\partial \Phi \over \partial t}=k_{0}J_0,\quad
\quad
i=0,\quad \eqno(3)$$
where
$$k=\frac{4\pi\sqrt{G}}
{c}\quad{\rm and}\quad k_0=8\pi\sqrt{G}.$$
These equations are first-order differential equations whose
solution can be found given the source terms.
Assuming the spacial homogeneity of the $\Phi$-field it
becomes independent of spatial
coordinates therefore
it holds $J_1= J_2= J_3=0.$
The source of the $\Phi$-field is its {\it
own} mass density associated with the field energy density, i.e. $\Phi^2/8\pi
c^2$, therefore, it holds
$J_0=k_0\Phi^2/8\pi
c^2$. $J_0$ is dependent only on time. The energy
density associated with the field
is
$${E_{\Phi}} ={\Phi^2 \over 8\pi}.$$
Since the current 4-vector in the everywhere local rest
frame has only one non-zero component, $J_0$,
 Eqs.(3) become
$$\nabla\Phi= 0$$
$$-{1\over c}{d \Phi \over d t}= {4\pi\sqrt {G}\Phi^2\over
8\pi c^2} ={\sqrt {G}\Phi^2 \over 2c^2}.\quad \eqno(4)$$
whose solution is
$$\Phi(t)= {c \over \sqrt {G} (t+t_0)},$$
where $t_0$ is the integration constant given by the boundary condition.
$\Lambda \propto 1/t^2$ has been considered by several authors with
different physical motivations, e.g. Lau \cite{L}
adopted the Dirac large-number
hypothesis of variable $G$, Kendo and Fukui \cite{EF} and others
operated in the context of
a modified Brans-Dicke theory etc.

In analogy with the electromagnetic
field, the quaternionic field acts on the moving "charged" objects with
the Lorenz-like force.
In \cite{MK} we supposed that the $\Phi$-field interacts
with all form of energy and matter and the
coupling constant k is from the dimensional reason equal to $\sqrt{G}$.
The "charge" of the $\Phi$-field for a point mass
$m_{0}$ is $\sqrt {G}m_{0}$.
Since the momentum of a moving particle is $p_i=m_0v_i,\quad i=1,2,3,0$,
its
current is given as $J_i= \sqrt{G} m_0v_i=\sqrt{G}p_i$. For the
Lorentz-like  force
acting on this particle in the $\Phi$-field we get
$$F_i=c^{-1}\sqrt{G}m\Phi v_i=  c^{-1}\sqrt{G}\Phi p_i.\quad\eqno(6)$$
In what follows we shall study the possible effect of the $\Phi$-field on
the moving bodies in solar and galactic conditions on the large time
scale.

\section{Force of the $\Phi$-field exerted on the moving bodies}

It is to be expected that the cosmical
$\Phi$-field manifests itself in the present-day
solar and galactic physical conditions only:
(i) at the large mass concentrations, (ii) at the
large velocities of massive
objects
(iii) during the large time and space scales.
For the sake of simplicity, we confine ourselves to the non-relativistic
case, i.e.
we suppose that $m=$ const. and $v\ll c$. Then Eq.(6) turns out to be
$$ m\dot v =c^{-1}\sqrt{G}\Phi mv. \quad \eqno(7)$$
Below, we present three possible effects of the quaternionic field in the
solar and galactic conditions:\\
(i) {\it The increase of the velocity of the moving bodies in
the $\Phi$-field}. Since $c^{-1}\sqrt{G}\Phi=1/t$ we get a
simple differential
equation $\dot v=\beta v $ where $\beta=1/t$
the solution of which is $v= Ct$.
A free moving object in the quaternionic field
is accelerated by a constant acceleration $C$. This
acceleration is due to the immense smallness of $\beta\approx 1/10^{18}$
in the present-day extremely
small. As is well-known for a given time instant the Hubble constant
$H$ is equal in
the whole Universe. Supposing $\beta|_{0}=H$,
the solution of equation $\dot
v= Hv$ becomes $v=Hr +C,$ where $C$ is an integration constant.
Setting $C=0$ we get a Hubble-like law $v=H r$.\\
(ii) {\it The increase of the kinetic energy of the moving bodies in the
$\Phi$-field.}
The gain of kinetic
energy of a moving body per time unit in the quaternionic field
if $(f_i\parallel v_i)$ is
         $$ {dE\over dt}=F_iv_i = c^{-1}
\sqrt {G} \beta mv^2 =2 \sqrt{G}c^{-1}\Phi E_{kin}=2\beta E_{kin}.
\quad\eqno(9)$$
Again, the increase of the kinetic energy of a moving object is extremely
small. However, for a rapid rotating dense body it may represent a
considerable value.
For example, a pulsar rotating around its axis with the angular
velocity $\omega$ having the moment of
insertion $I$. Its kinetic energy is $E_{kin}\approx I\omega^2$ and its
change
in the quaternionic field is $dE_{kin}/dt\approx \beta I\omega^2\approx
10^{32}$ which is a value only of some orders of magnitude smaller than
the energy output of a pulsar \cite{HA}.\\
(iii) {\it The change of the kinetic parameters of the gravitationally
bounded moving bodies.} This
can be best demonstrated by describing
the motion of the Earth around the Sun taking into
account Eq.(7). It holds
$$\vec F_1+\vec F_2=-\frac{GM_{\odot}m_{\oplus}}{r^2} \vec r,\quad \eqno(10)$$
where $\vec F_1=m_{\oplus}\vec {\ddot r}$ and $\vec F_2=
m_{\oplus}\vec {\beta \dot r}$.
Inserting $\vec F_1$ and $\vec F_2$ in
Eq.(10) we have
$$\frac{d}{dt}(\beta \dot {\vec r})=-\beta \frac{GM_{\odot}}{r^2} \vec r$$
from which it follows
$$\beta r^2 \dot \phi=const.=h\quad \eqno(11)$$
The reciprocal radius $u$ satisfies the equation
$$\frac{d^2 u}{d \phi} +u=\beta^2 \frac{GM_{\odot}}{h^2},$$
the general solution of which is
$$u(\phi)=\frac{C}{t^2}+c_2\cos(\phi)-c_1 \sin(\phi)$$
Setting $c_1=c_2=0$, i.e. supposing that the orbit is circle we get
$$r\sim K t^2\quad \eqno(12)$$
Hence, the distance of the Earth and the Sun varies with time like
$$r\sim \frac{1}{GM_{\odot}\beta^2}\sim \frac{1}{\beta^2}\sim t^2. \quad
\eqno(13)$$
According to Eq.(13) the distance between the Earth and the Sun is
increasing direct proportional to the square of time.
There have been several atomic time measurements of the period of the Moon
orbiting around the Earth. A description of the work of some
independent research groups can be found in Van Flander's article
\cite{FL}. We simply point that after subtracting the gravitational
perturbative (tidal) effects, Van Flanders gives
$$\frac{\dot P}{P}=\frac{\dot n}{n}=(3.2\pm 1). 10^{-10}/{\rm yr},$$
where $n=2\pi /P$ is the angular velocity. Using Eq.(11) we get
comparable value
$$|\dot n/n|\approx 4.10^{10}/{\rm yr}.$$
However, given the complexity of the
data analysis, we must certainly await further confirmation by
different, independent test before concluding whether the $\Phi$-field
really affects the motion of cosmic bodies. Nevertheless, it can be
asserted that at present, there exists no evidence against the influence
of the $\Phi$-field on the moving bodies at the level of the supposed
present-day intensity of the $\Phi$-field.
Due to large value of the cosmic time the present-day effects of
the $\Phi$-field lei on the limit of the observability. However, they
had, probably, strong influence in the early universe . For
example, the strong $\Phi$-field can destabilize the large rotating mass
concentration (e.g. quasars) forming from them the present-day galaxies.
We note that the enlargement of
distance of the Earth and the Sun is also suggested by the large numbers
hypothesis presented by Dirac in 1937 (for detail see \cite{CA}). In
this hypotheses Dirac supposed that $G\propto 1/t$. The astrophysical and
geological consequences of this hypothesis are discussed in details in
\cite{JO}\cite{WES}.

\section{The creation of particle  in the
$\Phi$-field}

Another interesting effect of the cosmic $\Phi$-field is the
possibility of the creation of real particles from the virtual ones.
Particle creation in nonstationary strong fields is well-known phenomenon
studied
intensively in seventies ( see, e.g.,\cite{GR}).
There are several proposed ways for the creation of real particles from
the virtual ones in the very strong and nonstationary gravitation field.
We propose here a new mechanism of the creation of real
particles from the virtual ones in the presence of the $\Phi$-field.\\
We note that our further
consideration on the creation of
matter from the vacuum quantum excitation are done by a semiclassic
way, although we realize that they should be
performed in terms of an adequate theory of
quantum gravity. However, as is well-known, when constructing a quantum
theory of gravity one meets conceptual and technical problems. The usual
concepts of field quantization cannot be simple applied to gravity
because standard field quantization (e.g., elm field) is normally done
in flat spacetime. It is impossible to separate the field equations and
the background curved space because the field equations determine the
curvature of spacetime. Moreover, the classical quantized field
equations are linear and that of gravitation are non-linear and
weak-field linearized gravitation field is not renormalizable.
There is even no exact criterion on which time and
space scales one has necessary to apply quantum laws for
gravitation field. Therefore, we take
the inequality $|A| \leq h $, where $A$
is the classical action, as a criterion for a possible application of
quantum physics in gravity.

According to
quantum theory, the vacuum contains many virtual particle-anti-particle
pairs whose lifetime
$\Delta t$ is bounded by the uncertainty relation $\Delta E\Delta t>h$
\cite{AI}
The proposed mechanism for the particle creation in the $\Phi$-field is
based on the force relation (6).
During the lifetime of the virtual particles
the Lorentz-like force (6) acts on
them and so they gain energy. To estimate this
energy we use simple heuristic arguments.
As is well-known, any virtual particle can only exist within limited
lifetime and its kinetics is bounded to the uncertainty relation
$\Delta p \Delta x>h$.
Therefore, the momentum of a virtual particle $p$ is approximately
given as $p\approx h\Delta x^{-1}$.
If we insert this momentum into Eq.(6) and multiply it by
$\Delta x$, then the energy of virtual
particle $\Delta E$, gained from the ambient $\Phi$-field during
its lifetime, is
$$F\Delta x= \Delta E =\sqrt{G} \Phi(t){h\over c}.\quad \eqno(15)$$
When the $\Phi$-field is sufficiently strong then it can supply enough
energy to the virtual particles during their lifetime and so
spontaneously create real particles
from the virtual pairs. The energy necessary for a particle to be created
is equal to $m_vc^2$ ($m_v$ is the rest mass of the real particle).
At least, this
energy must be supplied from the ambient $\Phi$-field to a virtual particle
during its lifetime. Inserting $\Phi$ into Eq. (15), we have
$$\Delta E\approx {h\over (t+t_0)},$$
Two cases may occur:
(i) If $m_vc^2<\Delta
E$, then the energy
supplied from the $\Phi$-field is sufficient for creating real particles of
mass $m_v$
and, eventually, gives them an additional kinetic energy.
(ii) If $m_v>c^2\Delta
E$, then
the supplied energy is not sufficient for creating the real particles  of
mass $m_v$ but only the energy excitations in vacuum.\\
The additional kinetic
energy of the created particles, when $\Delta E>m_vc^2$, is
$$E_{kin}=\Delta E -m_vc^2={h\over (t+t_0)} - m_vc^2.$$

As is well-known in Friedmann's cosmology the scale parameter $R$ varies
in time according to the
equation
$$\ddot R={4\pi G\over 3}[-\rho -3p +2\Lambda]R.\quad\eqno (18)$$
If we assume that at the very
beginning of the cosmical evolution no ordinary matter was presented
only the energy of the cosmic $\Phi$-field then we have
$$\ddot R\approx 2\Lambda R.$$
The solution to this equation describes the exponential expansion of
the early universe
$$R(t) \approx A\exp (2\Lambda).$$
This inflationary
phase of the cosmological evolution is supposed to stop when  an
massive creation of mass particles in the $\Phi$-field
began by means of mechanism described above.
The proposed cosmological evolution
started with purely
field-dominated epoch during which the inflation took place, after which
a massive creation of particles began. During the time interval
$(\approx 0,10^{-20})$, the masses of the created particles lei in the
range from $10^{-5}$ to $10^{-27}$ g.
Their kinetic energy
was $E_{kin}=[h/(t+t_0)]-m_0c^2$.
$E_{kin}$ of the created nucleons
has reached values up to $10^{-5}$ erg, which corresponds to
the temperature of $10^{21}$ K.
Today, energies of the virtual pairs, gained during their lifetime,
are immense small, therefore, they represent only a certain local energy
excitations of the vacuum.
\section{Conclusion}
From what has been said so far it follows:\\
(i) The $\Phi$-field exerts a force on the moving cosmic bodies
which is given by Eq.(6).\\
(ii) This force has negligible effects on moving bodies
in the present time but might be important
in the early stage of cosmological evolution.\\
(iii) A creation of real particle occurs
in the $\Phi$-field  which is
proportional to intensity of the $\Phi$-field.\\
Summing up, we can state that if the cosmic quaternionic field does exist then
it affects the kinetic parameters of moving bodies, causes the creation
of real particle from virtual one and enormously enlarges the temperature of the early universe.


\begin{thebibliography}{99}

\bibitem {ST}
M. S. Turner, Physics Today {\bf 56} (2003), 10, (April 2003).
%L. Wang and P. J. Steinhardt,  Astrophys. J. {\bf 508} (199) 483;
%I. Zlatev and P. J. Steinhardt, Phys.Lett {\bf B 459} (1999) 570;
%A. Albert and C. Skordis, Phys. Rev. Lett. {\bf 84} (2000) 2076.
%L. Wang, R, Caldwell, J. Ostriker and P. J. Steinhardt, Astrophys.
%J. {\bf 335} (2000) 17.
\bibitem{OO}
J. M. Overduin and F. I. Cooperstock, Phys. Rev. {\bf D 58} (1998) 43506.
\bibitem{EF}
M. Endo and T. Fukui, Gen. Relat. Gravit. {\bf 8} (1977) 833.
\bibitem{B}
O. Bertolami, Nuovo Cimento {\bf B 93} (1986) 36.
\bibitem{BS}
M. S. Berman and M. M. Som, Int. J. Theor. Phys. {\bf 29}  (1990) 1411.
\bibitem{L}
Y-K. Lau, Aust. J. Phys. {\bf 38}  (1985) 547.
\bibitem{BE}
A. Beesham, Gen. Relat. Grav. {\bf 26} (1994) 159.
\bibitem{LN}
J. L. Lopez and D. V. Nanopoulos, Mod. Phys. Lett. {\bf A 11} (1996) 1.
\bibitem{AL}
A. S. Al-Rawaf and M. O. Taha, Gen. Rel. Gravit. {\bf 28} (1995) 935;
Phys. Lett. {\bf B 366} (1996) 69.
\bibitem{ARB}
Arbab I. Arbab, arXiv:gr.-qc/9905066 vA  (14 May 2001); J. Cosmology and
Astroparticle Phys. {\bf 05} (2003) 008, 1.
\bibitem{A}
Anderson, R. and G. Ioshi (1993) Physics Essays {\bf 6}, 308.
\bibitem{MM}
Majern\'{\i}k, V., Advances in Applied Clifford Algebras.
{\bf 9}, 119-130 (1999).
\bibitem{SIG}
A. Singh, {\it Unified field theory based on new theory of gravitation
and the modified theory of electromagnetism.} (Virginia Polytechnic
Institute and state University: 1979);
Lett. Nuovo Cimento {\bf 33}, (1982), 457.
\bibitem{GR}
Grib, A. A.,  Mamayev, S. G. and Mostepanenko V. M. (1988) {\it Vacuum
quantum  effect in strong field}. Moscow, Energoatomizdat.
\bibitem{WES}
Wesson, P. S. (1978) {\it Cosmology and Geophysics.} A. Hilger, Bristol.
\bibitem{MK}
V. Majern\'{\i}k, Gen. Rel. Grav. {\bf 35} (2003), 1831 (in print).
\bibitem{CA}
V. M. Canuto, New Trends in Cosmology. Riv. Nuov. Cim. Ser.1, Vol.1,N.2,
1978.
\bibitem{FL}
Van Flandern, Astrophys. J. {\bf 248} (1981), 813.
\bibitem{JO}
P. Jordan, Schwerkraft und Weltall, Viewig, Braunschweig, 1954.
\bibitem{HA}
D. Hewish ter, Contemp. Phys. {\bf 16} (1975), 333.
\bibitem{AI}
I. J. R. Aitchinson, Contemp. Phys. {\bf 26} (1985), 333.
\end{thebibliography}
\end{document}